\def\cH{{\cal H}}
\def\cK{{\cal K}}
\def\cL{{\cal L}}
\def\cM{{\cal M}}
\def\ket#1{\left|\, #1\right\rangle}
\def\bra#1{\left\langle #1\right|}
\def\bracket#1#2{\left.\left\langle #1\right|#2\right\rangle}

\documentclass{article}
\usepackage{hyperref}
\usepackage{graphicx}
\title{Effective Quantum Time Travel}
\author{George Svetlichny\footnote{Departamento de Matem\'atica, Pontif\'{\i}cia Universidade Cat\'olica, Rio de Janeiro, Brazil \newline
svetlich@mat.puc-rio.br \hfill \url{http://www.mat.puc-rio.br/\~svetlich}}}
\begin{document}
\maketitle

\begin{abstract}
The quantum teleportation protocol can be used to probabilistically simulate a quantum circuit with backward-in-time connections. This allows us to analyze some conceptual problems of time travel in the context of physically realizable situations, to realize encrypted measurements of future states for which the decryption key becomes available only after the state is created, and to probabilistically realize a multistage quantum state processing within the time needed to complete only one stage. The probabilistic nature of the process resolves any paradox.
\end{abstract}

Consider that part of  the teleportation protocol for qubits which makes no use of classical communication, as shown in Fig. \ref{fig1}.
\begin{center}
\begin{figure}[hb]
\begin{center}
\begin{picture}(90,120)(0,0)
\put(0,80){\framebox(40,20)}
\put(20,90){\makebox(0,0){\({\cal M}\)}}
\put(50,0){\framebox(40,20)}
\put(60,20){\line(-1,2){30}}
\put(70,10){\makebox(0,0){\(\Psi _{00}\)}}
\put(0,60){\line(1,2){10}}
\put(80,20){\line(1,2){10}}
\put(5,55){\makebox(0,0){A}}
\put(85,45){\makebox(0,0){B}}
\qbezier[8](10,100)(10,110)(10,120)
\qbezier[8](30,100)(30,110)(30,120)
\put(6,72){\vector(1,2){0}}
\put(86,32){\vector(1,2){0}}
\put(44,52){\vector(-1,2){0}}
\end{picture}
\end{center}
\caption{}\label{fig1}
\end{figure}
\end{center}
Here \(\Psi _{00}=\frac1{\sqrt{2}}(\ket 0\ket 0+\ket 1\ket 1)\) is the totally entangled 2-qubit state, part of the Bell basis \(\Psi _{xy}=\frac1{\sqrt{2}}(\ket x\ket y+(-1)^y\ket{x+1}\ket{y+1})\) where \(x,\,y\) are qubit indices, that is \(0\) or \(1\), and all sums are modulo \(2\). \(\cM\) is a measurement in a basis containing \(\Psi _{00}\), such as the Bell basis itself. The dotted lines represent two classical bit of information which we'll ignore for the time being. There is a fundamental identity in Hilbert space theory (I consider only finite dimensions): \(\cL(\cH_1\otimes \cH_2, \cH_3)\simeq \cL(\cH_1, \cH_3\otimes \cH_2^*)\) where by \(\cL(\cH,\cK)\) I denote the space of linear maps from \(\cH\) to \(\cK\) and by \(\cH^*\) I denote the dual space to \(\cH\), the space of bras. I identify \(\cH^{**}\) with \(\cH\) itself. Thus we can identify any entangled state \(\Phi =\sum c_i\ket {\alpha _i}\otimes \ket {\beta _i}\in \cH\otimes \cK\) with a map \(F_{\Phi }:\cH^*\to \cK\) by \(\bra \gamma \mapsto \sum c_i\bracket{\gamma }{ \alpha _i}\ket {\beta _i}\), a fact behind the many equivalences between states and channels mentioned in the literature. Likewise any bra of an entangled state \(\Psi^* =\sum d_i\bra {\alpha _i}\otimes \bra {\beta _i}\in (\cH\otimes \cK)^*\simeq \cH^*\otimes \cK^*\) can be identified with a map \(G_{\Psi^* }:\cH\to \cK^*\) by \(\ket \gamma \mapsto \sum d_i\bracket{ \alpha _i}{\gamma} \otimes \bra {\beta _i}\). Consider now those measurement incidents of \(\cM\) in which the state is projected onto the \(\Psi _{00}\) subspace. This projector is \(\ket{\Psi _{00}}\bra{\Psi _{00}}\) so the action of the bra of this projector can be considered as a map from \(\cH\) to \(\cH*\) (the qubit Hilbert space) which is then transformed by the ket \(\Psi _{00}\) in the lower box again to an element of \(\cH\). A simple calculation shows \(F_{\Psi _{00}}\circ G_{\Psi _{00}^*}=\frac 12 I\), so aside from an overall scalar the qubit at point B of the diagram is the same as at point A. When the measurement projects onto another Bell state, the qubit at point B is a unitary transform of the one at point A, and this fact makes deterministic teleportation work as then {\em after the result of the measurement is known\/} one can transform the qubit issuing from B to be identical to the one entering A \footnote{I learned of this view of teleportation at the 2004 meeting of the International Quantum Structures Association in Denver during a talk by Bob Coecke. See arXiv:quant-ph/0402014.}. What we have in the case of projection onto \(\Psi _{00}\) is {\em effective\/} time travel. After the fact of the measurement \(\cM\) taken place there is no empirical way to falsify the statement that the qubit at A did travel back in time to B, but this is not true time travel. By true time travel I mean one whose denial can be falsified by empirical evidence. One can however ask if any of the supposed effects and benefits of  supposed true time travel do somehow exist in this case. The surprising answer is yes, but obviously those that cannot lead to time travel paradoxes.  In these cases, time travel is a {\em reading\/}  of the situation which can otherwise be analyzed in usual quantum mechanical terms, but this reading makes these process conceptually easier to follow and so I'll use this metaphor. Thus I will use expressions such as ``send back in time'' without the quotes, knowing I am dealing in the restricted context of this paper. I consider only multipartite qubit systems as extensions to higher dimensions would be straightforward.

When the measurement result of \(\cM\) is not projection onto \(\Psi _{00}\) but onto another stated \(\Psi \) then the qubit at B is the \(F_{\Psi}\circ G_{\Psi _{00}^*}\) transform of the one at A.  One can still describe  this with a time travel metaphor in which time travel itself changes the input state  to another one in a linear way. One then comes out in the past transformed from what one was upon entering the time machine.

Consider now a qubit quantum circuit with one backward-in-time connection as in Fig \ref{fig2}, which may be considered as 
\begin{figure}[h]
\begin{center}
\begin{picture}(80,110)(0,0)
\put(0,60){\framebox(30,20)}
\put(15,70){\makebox(0,0){\(V\)}}
\put(50,30){\framebox(30,20)}
\put(65,40){\makebox(0,0){\(W\)}}
\qbezier(20,80)(20,95)(30,95)
\qbezier(30,95)(40,95)(40,80)
\put(40,30){\line(0,1){50}}
\qbezier(40,30)(40,15)(50,15)
\qbezier(50,15)(60,15)(60,30)
\put(10,0){\line(0,1){60}}
\put(20,0){\line(0,1){60}}
\put(10,80){\line(0,1){30}}
\put(70,0){\line(0,1){30}}
\put(60,50){\line(0,1){60}}
\put(70,50){\line(0,1){60}}
\end{picture}
\end{center}
\caption{}\label{fig2}
\end{figure}
part of a larger circuit with no other backward-in-time connections.  If instead of making the connection we kept the output and input as such, we would have a multipartite unitary map. What the circuit with the connection represents is then a partial trace of this map preceded or followed (in this case) by a swap map (which is unitary) of the two Hilbert spaces involved. We are thus led to consider the situation represented by Fig. \ref{fig3}
\begin{figure}[h]
\begin{center}
\begin{picture}(40,90)(0,0)
\put(0,30){\framebox(30,30)}
\put(15,45){\makebox(0,0){\(U\)}}
\qbezier(20,60)(20,75)(30,75)
\qbezier(30,75)(40,75)(40,60)
\put(40,30){\line(0,1){30}}
\qbezier(40,30)(40,15)(30,15)
\qbezier(30,15)(20,15)(20,30)
\put(10,0){\line(0,1){30}}
\put(10,60){\line(0,1){30}}
\end{picture}
\end{center}
\caption{}\label{fig3}
\end{figure}
with \(U\) unitary and  where the loop is a qubit Hilbert space, and the other lines represent the tensor product of all the other qubit spaces in the circuit. The loop is a quantum mechanical analog of a closed time-like curve (CTC). We now modify this figure by the scheme of Fig. \ref{fig1} as displayed in Fig. \ref{fig4}.
\begin{figure}[h]
\begin{center}
\begin{picture}(70,130)(0,0)
\put(0,40){\framebox(30,30)}
\put(15,55){\makebox(0,0){\(U\)}}
\put(10,0){\line(0,1){40}}
\put(10,70){\line(0,1){60}}
\put(30,90){\framebox(40,20)}
\put(30,0){\framebox(40,20)}
\put(50,100){\makebox(0,0){\({\cal M}\)}}
\put(50,10){\makebox(0,0){\(\Psi _{(00)}\)}}
\put(20,70){\line(1,1){20}}
\put(20,40){\line(1,-1){20}}
\put(60,20){\line(0,1){70}}
\qbezier[8](40,110)(40,120)(40,130)
\qbezier[8](60,110)(60,120)(60,130)
\end{picture}
\end{center}
\caption{}\label{fig4}
\end{figure}
Each time the measurement projects onto \(\Psi _{00}\), the arrangement in Fig. \ref{fig4} acts, up to an overall scalar multiple, exactly as Fig. \ref{fig3}. For the sake of greater clarity I present an algebraic proof of this fact. Let \(U_{ap}^{bq}\) be the matrix of \(U\) where input indices are lower and output upper. Indices \(p\)  and \(q\) refer to the qubit Hilbert space in the computation basis. The partial trace is \(U_{ap}^{bp}\) using the summation convention that repeated indices are summed over. For Fig. \ref{fig4} the Bell state \(\Psi _{00}\) corresponds to a matrix \(\Psi ^{rs}\) and a projection on this state to a matrix \(\Psi _{ij}\Psi ^{rs}\). So  Fig. \ref{fig4} is \(U_{ap}^{bi}\Psi _{ij}\Psi ^{rs}\Psi^{jq} \). Now in the computation basis, \(\Psi ^{rs}=\frac1{\sqrt 2}\delta ^{rs}\) and  \(\Psi _{ij}=\frac1{\sqrt 2}\delta _{ij}\) so Fig. \ref{fig4} is precisely \(\frac12U_{ap}^{bp}\Psi ^{rs}\) and so aside from the factor \(1/2\) and the disjoint state \(\Psi ^{rs}\) one has precisely the partial trace.  We are still not making use of the classical information obtained by the measurement. The usual teleportation scheme is a particular case of this where \(U\) is the swap map. The partial trace of the swap map is \(I\) and one can use the classical information to achieve an identity transform, up to a factor of \(\frac12\), for each of the possible measurement outcomes.

An instructive example of the above construct is when \(U\) is the CNOT gate, whose action is \(\ket x\ket y\mapsto \ket x\ket{x+y}\). The modified partial trace of this is given by Fig. \ref{fig5}.

\begin{figure}[h]
\begin{center}
\begin{picture}(70,140)(0,0)
\put(30,100){\framebox(40,20)}
\put(30,0){\framebox(40,20)}
\put(50,110){\makebox(0,0){\({\cal M}\)}}
\put(50,10){\makebox(0,0){\(\Psi _{00}\)}}
\put(20,60){\line(1,2){20}}
\put(20,60){\line(1,-2){20}}
\put(60,20){\line(0,1){80}}
\qbezier[8](40,120)(40,130)(40,140)
\qbezier[8](60,120)(60,130)(60,140)
\put(20,60){\makebox(0,0){\(\otimes\)}}
\put(0,60){\circle*{4}}
\put(0,0){\line(0,1){140}}
\put(0,60){\line(1,0){17}}
\end{picture}
\caption{}\label{fig5}
\end{center}
\end{figure}

The partial trace of the CNOT gate is twice the orthogonal projection onto \(\ket 0\). The modified partial trace with \({\cal M}\) a Bell basis measurement is actually a projective measurement in the computation basis. Assume the state on the control line is \(\phi =\alpha\ket 0+\beta \ket 1\) , then after the creation of \(\Psi _{00}\) and before the gate acts, the state is \(\phi \otimes\Psi _{00}\). After the gate acts the state is \(\alpha \ket 0\otimes\Psi _{00}+\beta \ket1\otimes\Psi _{10}\). So in the subsequent measurement projections onto the Bell states \(\Psi _{01}\) and \(\Psi _{11}\) never occur and the two that do occur correspond to projecting \(\phi \) onto \(\ket 0\) or \(\ket 1\) respectively with  probabilities according to projective measurement in the computation basis. The partial trace of CNOT has been considered an analog of a paradoxal time-travel situation \cite{deut:PRD44.3197} where an object goes back in time and changes its state in a contradictory way. Thus in the case the control qubit is \(\ket 1\), if the state \(\ket 1\) leaves the gate on the right and loops back in time, then upon arriving at the gate it must change and leave as  \(\ket 0\), a contradiction. This contradiction manifests itself formally by the fact that the partial trace, being twice the orthogonal projection onto \(\ket 0\), annihilates \(\ket 1\). Of course one cannot physically annihilate a state, an operator annihilating a state is physically realized only in the company of others, such as in a POVM, where the transformed state is given by another operator acting on the original. This is what happens in the modified partial trace. If we dramatically call \(\ket 1\) ``alive'' and \(\ket 0\) ``dead'', then with \(\ket 1\) at the control, the gate kills but also resurrects. The only state that can loop around consistently is the Schr\"odinger cat state \(\ket 1+\ket 0\)\footnote{One may also want to admit \(\ket 1-\ket 0\) since going around the loop it changes only by phase, but this is inconsistent with the action of the gate as a vector transformer and not a ray transformer.}. This is similar to fixed-point versions of consistent time travel \cite{kuta:PhilSci70.1098}. Of course in a partial trace, it's not a particular state that loops around, but a whole (no matter which) orthonormal basis; this is what computing a trace means. Nevertheless, such traces do constitute  metaphors for CTC's and analyzing them as such provides a few insights into what time means in quantum situations. Consider a hapless physicist who has to traverse the loop of  the partial trace of CNOT. Leaving the gate alive he has to, logically,  be leaving it dead and vice versa. How does Fig. \ref{fig5} resolve the paradox? Assume the control state is \(\phi =\alpha\ket 0+\beta \ket 1\) as above. If the exit state on the control line is \(\ket 0\), this can be construed that the gate operated with \(\ket 0\) on the control and also that time travel succeeded as the measurement projects onto \(\Psi _{00}\). The gate mercifully does not kill the traveler. If the exit state on the control line is \(\ket 1\) this can be construed that the gate operated with \(\ket 1\) at the control. The measurement projected onto the \(\Psi _{10}\) Bell state. The composition \(F_{\Psi _{00}}\circ G_{\Psi {10}^*}\) acts by \(\ket x\mapsto \frac 12\ket {x+1}\) so time travel itself kills our physicist but the gate mercifully resurrects him. No paradox here and there couldn't be one as Fig. \ref{fig5} is a physically realizable construct. Deutch in Ref. \cite{deut:PRD44.3197} had to appeal to density matrices to resolve time travel paradoxes as contradictions remained with individual states. This does not happen in our case.

Of course the above is a {\em narrative\/} and should not be simply accepted as a description of a physical process. All philosophically or scientifically motivated discussions of time travel have been likewise narratives as the time travel process, or time machine, is necessarily merely {\em hypothesized\/} to have certain properties not being able to refer to physically realizable situations.  This type of narrative is at best a meta-theoretic discussion, for instance exploring the question as to whether  the existence of supposed true time travel is consistent with present physical laws or do these need to be modified to admit it. In our narratives neither the claim nor the denial of time travel can be empirically falsified. This non-falsifiability makes both the claim and the denial non-scientific assertions. On the other hand our narrative takes place in the context of a physically realizable process and so cannot lead to any contradiction. We thus have a source of time travel narratives in which all paradoxes are resolved and this has interesting philosophical and scientific implications.

Deutch \cite{deut:PRD44.3197, deut:SciAm031994.68} also proposed a resolution of time-travel paradoxes via Everett's many-worlds interpretation. In short, a time-traveler going into the past enters a parallel universe and so, even if she prevents her younger self from entering the time machine, this does not cause a paradox as her  younger self is one in a parallel universe. There is an Everettian split occurring in time-travel and one never travels back to one's own past. If we interpret our narrative in Everettian terms, the split occurs upon measurement \(\cM\), that is, upon {\em entering\/} the time machine. The operators of the time machine (whoever is making the measurement \(\cM\)) know if the traveler went back unscathed or transformed. For consistency however, since the traveler going to the past is either unscathed or transformed according to the measurement outcome, one has to assume the worlds were split in the past also. Thus to admit our time travel narrative along with the Everett interpretation one must conclude that the world does not bifurcate with measurements, but that measurements are nexus where parallel worlds meet but are split both in the past and the future. This is a time-symmetric version of Everett and seemingly more easily visualized in relativistic contexts.

Leaving narratives aside, I now explore some possible practical applications of effective time travel. Suppose we have a bipartite qubit state with possibly separated parts and wish to perform a joint measurement on them. There are two obvious ways to proceed if the parts are separated. One can physically  route one of the qubits to the location of the other (or both to a common location) via a physical medium, or teleport one of them to the location of the other (or both to a common location), and then perform a joint measurement. There is a third way, send both qubits back in time to a common location and perform a joint measurement there and then. This is exemplified by Fig. \ref{fig6}.
\begin{figure}[h]
\begin{center}
\begin{picture}(180,120)(0,0)
\put(0,80){\framebox(40,20)}
\put(20,90){\makebox(0,0){\({\cal M}_1\)}}
\put(40,0){\framebox(40,20)}
\put(50,20){\line(-1,3){20}}
\put(60,10){\makebox(0,0){\(\Psi _{00}\)}}
\put(0,60){\line(1,2){10}}
\put(70,20){\line(1,1){10}}
\put(5,55){\makebox(0,0){A}}
\put(70,30){\framebox(40,20)}
\put(100,0){\framebox(40,20)}
\put(110,20){\line(-1,1){10}}
\put(130,20){\line(1,3){20}}
\put(120,10){\makebox(0,0){\(\Psi _{00}\)}}
\put(140,80){\framebox(40,20)}
\put(180,60){\line(-1,2){10}}
\put(177,55){\makebox(0,0){B}}
\put(90,40){\makebox(0,0){\({\cal M}_0\)}}
\put(160,90){\makebox(0,0){\({\cal M}_2\)}}
\qbezier[8](10,100)(10,110)(10,120)
\qbezier[8](30,100)(30,110)(30,120)
\qbezier[8](150,100)(150,110)(150,120)
\qbezier[8](170,100)(170,110)(170,120)
\qbezier[8](80,50)(80,60)(80,70)
\qbezier[8](100,50)(100,60)(100,70)
\end{picture}
\end{center}
\caption{}\label{fig6}
\end{figure}
Here \({\cal M}_1\) and  \({\cal M}_2\) are measurements in a basis containing \(\Psi _{00}\) and \({\cal M}_0\) the joint measurement we wish to perform on the bipartite state \(\Phi _{\rm AB}\) with qubits at locations A and B. For ease of expression I shall refer to the time of measurement \({\cal M}_0\) as ``today'' and that of measurements \({\cal M}_1\) and \({\cal M}_2\) as ``tomorrow''. Whenever the measurements tomorrow project onto \(\Psi _{00}\) at both locations the result of measurement \({\cal M}_0\) today can be considered a legitimate measurement of tomorrow's state  \(\Phi _{\rm AB}\). Of course this is a probabilistic situation; if we perform a large run of say \(N\) measurements today with the  corresponding  run of pairs of measurements tomorrow, only a fraction of the measurements today can be considered legitimate measurements of tomorrow's state, those that result tomorrow in projection onto \(\Psi _{00}\) in both measurements. We have today thus {\em probabilistic\/} information about tomorrow. This in itself is not at all  unusual. I could write down today all the possible outcomes of tomorrow's lottery drawing,  one of these is correct with known probability, and I will only know tomorrow which one this is when the lottery is drawn. Today's information is useless. The quantum mechanical situation goes beyond this. This is most readily seen when all three measurements in Fig. \ref{fig6} are measurements in the Bell basis. One has for any Bell state \(\Psi _{xy}\) that, up to an overall phase,  \(G_{\Psi ^*_{xy}}=G_{\Psi ^*_{00}}\circ\sigma _{xy} \) where the \(\sigma _{xy}\) are appropriately labeled unitary (and hermitian) sigma matrices with \(\sigma _{00}\) the identity \footnote{Specifically \(\sigma _{10}=\sigma _x\), \(\sigma _{01}=\sigma _y\), and \(\sigma _{11}=\sigma_z\) where the right-hand sides are the conventionally named sigma matrices.}. Thus whenever a measurement tomorrow projects onto \(\Psi _{xy}\), the qubit arriving today is transformed by \(\sigma _{xy}\). Thus in all cases the state measured today is \(\sigma _{xy}\otimes \sigma_ {uv}\Phi _{\rm AB}\). The probability of projecting this state onto a Bell state \(\Psi _{wz}\) is the same as projecting \(\Phi _{\rm AB}\) onto \(\sigma _{xy}\otimes \sigma_ {uv}\Psi _{wz}\) {\em which is another Bell state\/} \(\Psi _{w'z'}\). Thus reassigning today's outcome of projection onto \(\Psi _{wz}\) to a projection onto \(\Psi _{w'z'}\), which we can do tomorrow, we once again have a legitimate measurement of \(\Phi _{\rm AB}\). The \(N\) measurements of today's run is thus a {\em ciphertext\/} of a run of measurements of tomorrow's state. The \(N\) pairs of tomorrow's measurements is the key necessary to decipher today's results. This is different from the lottery case. Today's measurement are useless by themselves (just as in the lottery case) since they are encrypted, but tomorrow's measurements are also useless by themselves since they are local measurements from which one cannot deduce the run of joint measurements. This splitting of information between today and tomorrow is made possible by quantum entanglement and does not seem to have a classical counterpart.

One may feel there is a paradox here as one could between today and tomorrow, after all the \(N\) measurements today are performed, change one's mind about what state to create tomorrow. How is it that today's measurements still form a legitimate run of measurements of tomorrow's state? What changes is the key needed to decipher. This is similar to the situation of receiving a text consisting of a random sequence of letters and punctuation marks, like one that can  be created by a one-time pad. For such a ciphertext, {\em any\/} text of the same length could be the plaintext. Receiving different keys, one reads different plaintexts. For this to work classically, the sender of the key needs to know the ciphertext, in the quantum case, the key is generated without such knowledge.

Instead of the measurement \({\cal M}_0\) in Fig. \ref{fig6}, one can place a unitary gate. This allows us to probabilistically subject tomorrow's state to a unitary transformation today. But then we can go on, and instead of subjecting the outcome of this gate to another one tomorrow, perform this subsequent transformation also today. Thus any finte sequence of unitary transformations to be performed tomorrow, the day after tomorrow, the day after that, and so on, can all be probabilistically performed  today, and tomorrow one can find out if the endeavor succeeded. There may be some practical advantage to this, even paying the price of probability. Each transformation is performed on freshly created states, possibly thus lessening effects of decoherence. Whether one can also use the instances of tomorrow's measurements not projecting onto \(\Psi _{00}\) to some advantage, as in the joint measurement case, has not been investigated.

One can wonder if the above described processes occur in nature. At first thought this could only occur in complex highly organized system. Biological systems, especially neurobiological ones, come to mind, but this is pure speculation.

This research received partial financial support from the Conselho Nacional de Desenvolvimento Cient\'{\i}fico e Tecnol\'ogico (CNPq), and the Funda\c{c}\~ao de Amparo \`a Pesquisa do Estado do Rio de Janeiro  (FAPERJ).

\section*{Acknowledgements}This research received partial financial support from the Conselho Nacional de Desenvolvimento Cient\'{\i}fico e Tecnol\'ogico (CNPq), and the Funda\c{c}\~ao de Amparo \`a Pesquisa do Estado do Rio de Janeiro  (FAPERJ).

\end{document}